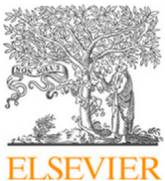
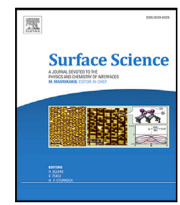

# Evolution of the surface morphology of GaSb epitaxial layers deposited by molecular beam epitaxy (MBE) on GaAs (100) substrates

Dawid Jarosz [a], Ewa Bobko [a,*], Marcin Stachowicz [b], Ewa Przeździecka [b], Piotr Krzemiński [a], Marta Ruszała [a], Anna Juś [a], Małgorzata Trzyna-Sowa [a], Kinga Maś [a], Renata Wojnarowska-Nowak [a], Oskar Nowak [a], Daria Gudyka [a], Brajan Tabor [a], Michał Marchewka [a]

[a] *Institute of Materials Engineering, Center for Microelectronics and Nanotechnology, University of Rzeszow, Al. Rejtana 16, 35-959 Rzeszow, Poland*
[b] *Institute of Physics, Polish Academy of Sciences, Al. Lotników 32/46 PL-02-668 Warsaw, Poland*

ABSTRACT

This study presents a demonstration of the surface morphology behavior of gallium antimonide (GaSb) layers deposited on gallium arsenide (GaAs) (100) substrates using three different methods: metamorphic, interfacial misfit (IMF) matrix, and a method based on a Polish patent application number P.443805. The first two methods are commonly used, while the third differs in the sequence of successive steps and the presence of Be doping at the initial growth stage. By comparing GaSb layers made by these methods for the same growth parameters, the most favorable procedure for forming a GaSb buffer layer is selected. Using GaAs substrates with a GaSb buffer layer is a cheaper alternative to using GaSb substrates in infrared detector structures based on II-type superlattices T2SL, such as InAs/GaSb. The quality of the GaSb buffer layer determines the quality of the subsequent layers that form the entire T2SL and affects factors such as dark current in terms of application.

## 1. Introduction

In photodetector technology based on type II InAs/GaSb superlattices operating in the infrared range, InAs/AlSb superlattices often serve as a barrier region [1,2]. Typically, these structures are grown on GaSb substrates (100) matched to the type II superlattice through interface engineering [3,4]. However, using cheaper GaAs substrates with a GaSb layer offers advantages due to their low IR absorption and favorable thermal properties [5]. The challenge is the significant lattice mismatch of 7.8% between GaAs substrates and GaSb layers [6]. MBE technology provides various ways to create GaSb buffer layers, with a GaSb layer deposited on a GaAs substrate mimicking a GaSb substrate. In this study, we compared the surface morphology behavior of GaSb layers using two commonly employed growth methods: metamorphic and interfacial misfit (IMF) matrix. We also investigated a method based on Polish patent application number P.443805, developed by our research group [7]. The samples were characterized using atomic force microscopy (AFM), high-resolution X-ray diffractometry (HD-XRD), scanning electron microscopy (SEM), and Nomarski optical microscopy. To determine the timing of GaSb layer solidification, we used an innovative approach: pyrometric measurement of the temperature of the (GaSb layer + GaAs substrate) system.

## 2. Experimental and methods

In the course of this study, we prepared three sets of samples utilizing distinct techniques to examine their behavior. Prior to depositing heteroepitaxial GaSb layers on quarter-cut two-inch wafers made of GaAs:Un (100) substrates using the metamorphic method, interfacial mismatch matrix method, and our patented technique described in Polish patent application number P.443805, we carried out numerous homoepitaxial processes for GaAs and GaSb to determine ideal growth parameters. All homoepitaxial GaAs and GaSb layers were grown at substrate temperatures of $T_S$ = 590 °C and $T_S$ = 530 °C, respectively, as indicated by the IRCON pyrometer. The calibration was performed separately for GaAs and GaSb substrates by observing the RHEED pattern during desorption of oxides from the substrate surface at $T_S$ = 580 °C and $T_S$ = 566 °C, respectively. For heteroepitaxial structures grown on GaAs:Un (100) substrates, a refresh layer of 200 nm thick GaAs with optimal growth parameters such as a V/III flux ratio of 16.78 for an approximately 948 nm/h growth rate and a substrate temperature of $T_S$ = 590 °C was first deposited. The subsequent deposition of the GaSb buffer layers utilized three distinct techniques with optimal conditions obtained for GaSb homoepitaxial growth, including a V/III flux ratio of 7.29 for an approximately 399 nm/h growth rate and a substrate temperature of $T_S$ = 530 °C. All sample sets were deposited using a solid source Riber Compact 21T (III-V) MBE system outfitted with standard ABN60 dual-zone effusion cells for In, Al, Ga, and valved arsenic and antimonic cracker VAC 500 and VCOR 300, respectively. The rotation speed during growth was set to 10 rpm (rotations per minute), and the substrate temperature $T_S$ ramp rate was 10 °C/min during the heating process and 20 °C/min during the cooling process. Additionally, we reported the temperature of the primary heater acquired from the thermocouple in this investigation and labeled it $T_H$. The characterization of all samples was carried out using various instruments such as Atomic Force Microscopy INNOVA BRUKER, Scanning Electron Microscope HELIOS NANOLAB650, Optical Microscope with Differential interference contrast (DIC) OLYMPUS DSX1000, High-Resolution X-ray Diffractometer EMPYREAN 3.

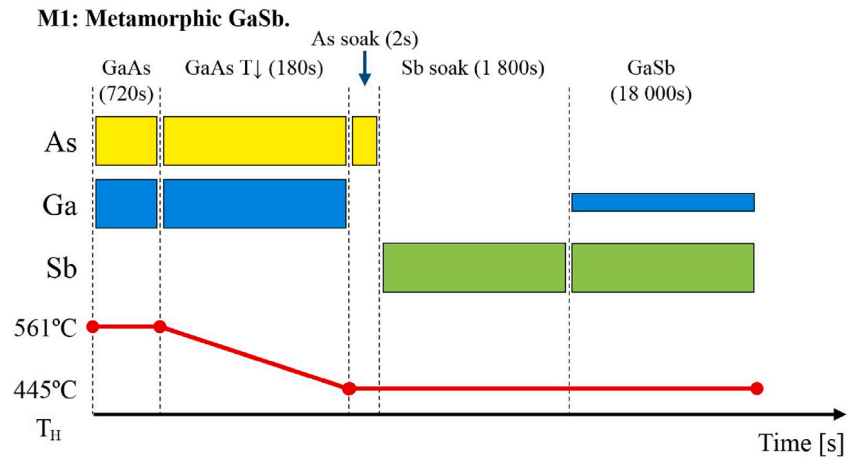

**Fig. 1.** The shutters sequences of MBE process during the growth of the GaSb metamorphic buffer layer described as method M1. The red line represents the temperature of the main heater $T_H$ and not the substrate $T_S$.

**M1:** Refers to the deposition of a GaSb buffer as the metamorphic layer [6]. In the first step, after removing surface oxides from the GaAs substrate, a 200 nm thick GaAs buffer layer was deposited under optimal growth conditions for homoepitaxial GaAs layers, including gallium flux, arsenic flux, and a substrate temperature of $T_S$ = 590 °C [7]. At the end of the GaAs buffer layer growth, the substrate temperature was decreased to $T_S$ = 510 °C, and the supply of gallium was terminated. Subsequently, the arsenic flux was replaced with an optimal antimony flux for the growth of homoepitaxial GaSb layers. Finally, a gallium flux with an ideal value for the growth of homoepitaxial GaSb layers was introduced to initiate the deposition of the GaSb layer [6].

**M2:** Buffer GaSb refers to the growth of a GaSb layer using an interfacial misfit (IMF) matrix. In the first step, following the deoxidation of the GaAs substrate, a 200 nm thick GaAs buffer layer was deposited with optimized growth conditions for homoepitaxial GaAs layers, as previously carried out in method M1. The GaAs buffer layer growth was then interrupted by reducing and cutting off both gallium and arsenic fluxes for 20 s. Next, an antimony flux of the optimum value for homoepitaxial GaSb layers was applied to wet the GaAs buffer layer. The substrate temperature was then lowered to $T_S$ = 510 °C. In the final step, a gallium flux with an optimal value for the growth of homoepitaxial GaSb layers was introduced to initiate the deposition of a GaSb layer [8].

**M3:** Buffer GaSb according to Polish patent application P.443805 [7]. It involves a process similar to methods M1 and M2, starting with the deposition of a 200 nm thick GaAs buffer layer at an approximate growth rate of 1 µm/h controlled by gallium flux. The growth rate was then reduced to approximately 120 nm/h by proportionally decreasing both the gallium and arsenic fluxes while maintaining a constant ratio of V/III elements at around 16.78. Afterward, the substrate temperature was lowered from $T_S$ = 590 °C without interrupting GaAs layer growth for 3 min to an optimal value for GaSb layer growth equal to $T_S$ = 510 °C and growth process interruption after cutting off gallium flux. Next, the gallium flux was interrupted, and the GaAs layer's surface was wetted with arsenic flux for one second followed by antimony flux for one minute before growing GaSb layers. The V/III flux ratio for GaSb layers was approximately 7.29. The procedure continued with the deposition of a Be doped GaSb layer with a linear falloff in Be concentration from 1e19 cm$^{-3}$ to 1e18 cm$^{-3}$ for approximately 400 nm, limited by gallium flux growth rate of approximately 400 nm/h. Then an approximately 40 nm thick Be doped GaSb layer with an exponential decay in Be concentration from 1e18 cm$^{-3}$ to 1e17 cm$^{-3}$ was deposited. Lastly, a pure 1600 nm thick GaSb layer was grown under optimized growth conditions. The change of optimized substrate temperature from $T_S$ = 510 °C to $T_S$ = 530 °C was notified after conversion group V from As to Sb flux. The pyrometer temperature indication has changed despite of maintained constant temperature at the main heater $T_H$ = 445 °C.

Figs. 1–3 show a schematic representation of the shuttering sequence for the procedures labeled M1, M2, M3 and the temperature of the main heater throughout the transition from GaAs to GaSb material.

## 3. Results and discussion

*3.1. Evolution of the surface morphology of GaSb epitaxial layers deposited on GaAs (100) substrates by a metamorphic method labeled M1, M2 and M3*

Buffer layers designed to reduce dislocation density on a mismatched substrate are called metamorphic buffers [9–11]. As the thickness of the GaSb buffer increases, dislocation density decreases [12]. IMF [13–16] allows for reduced dislocation density at the surface of the (110) buffer layer by propagating most dislocations at a 90-degree angle to the substrate/layer interface by converting As anions into Sb [15]. Be doping of GaSb buffer layers at the initial stage of crystallization (as in method M3) has also been observed to have a beneficial effect [8]. This study presents results from surface morphology measurements during GaSb buffer layer formation using methods M1, M2, and M3. Each section contains seven processes performed with identical growth parameters but varying GaSb buffer layer growth times. The first processes involved refreshing the GaAs substrate with a 200 nm GaAs layer cooled in the Sb flux. Subsequent processes had GaSb buffer layers grown by: 2.5, 5, 15, 25, 60, and 300 min. Figs. 4, 6, 8 present AFM measurements from each process, divided into three groups with a selected color scale for each group. The lower parts of these figures present plots of the substrate temperature $T_S$ read from the pyrometer during GaSb buffer layer growth with an exponential fit for non-random measurement points. Insets provide a comparison of GaSb layer growth time with surface roughness RMS from AFM for the (10 × 10) μm region. Figs. 5, 7, 9 contain SEM images from a cross-section of GaSb layers made according to methods labeled M1, M2, and M3 for selected GaSb layer aforementioned growth times. During growth, the substrate temperature increased from $T_S$ = 510 °C to $T_S$ = 532 °C due to exposure to thermal radiation from the cracker area of the Sb cell maintained at 900 °C, while the As cracker was set at 700 °C. The initiation of GaSb layer growth resulted in a rapid increase in substrate temperature reading that oscillated randomly between $T_S$ = 550 °C to $T_S$ = 560 °C for about 15 to 30 min of growth depending on the GaSb buffer growth method, before finally stabilizing at $T_S$ = 562 °C. The random behavior of the substrate temperature was attributed to the developing surface morphology of the crystallization front while the increase in substrate temperature is influenced by various factors, including exposure to thermal radiation from a Ga cell maintained at 1041 °C, changes in emissivity of the substrate associated with deposition of the GaSb layer, and energy yield associated with the crystallization process.

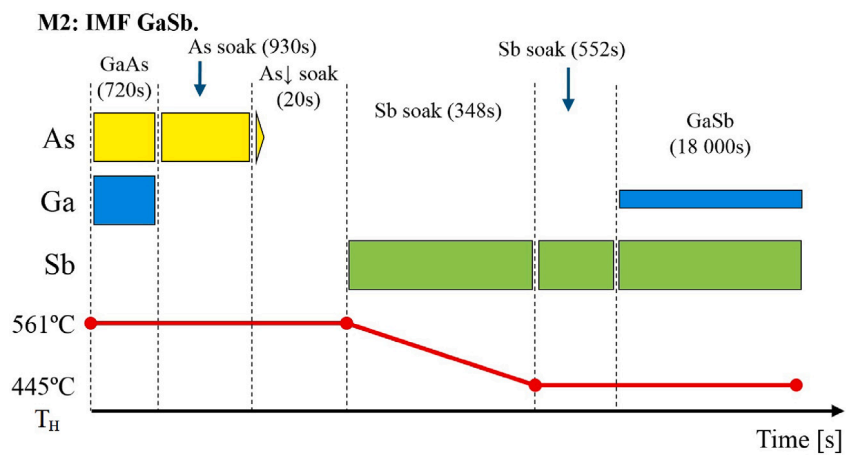

**Fig. 2.** The shutters sequences of MBE process during GaSb buffer layer growth using interfacial misfit (IMF) matrix method labeled M2. The red line represents the temperature of the main heater $T_H$ and not the substrate $T_S$.

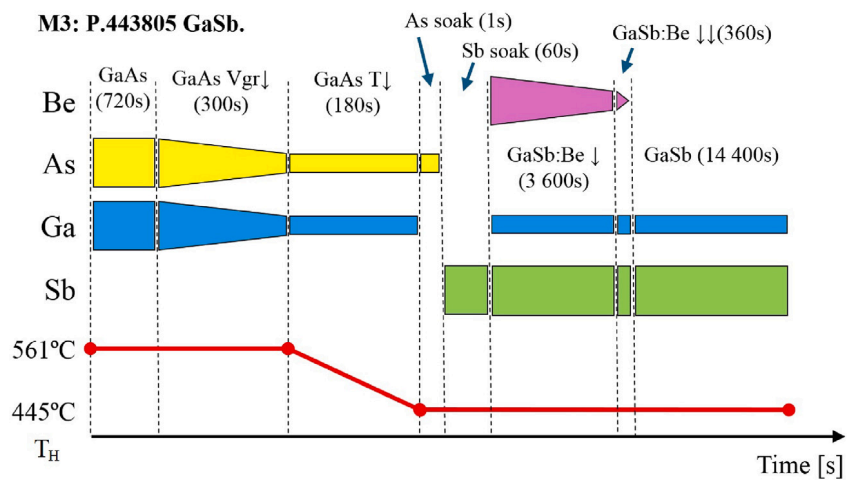

**Fig. 3.** The shutters sequences of MBE process during GaSb buffer layer growth using our method according to the Polish patent application P.443805 designated as M3. The red line represents the temperature of the main heater $T_H$ and not the substrate $T_S$.

*3.1.1. GaSb buffer layers deposited on GaAs (100) substrates by method labeled M1*

Due to wetting and cooling for about 30 min with Sb flux, the surface of the GaAs layer achieved an RMS surface roughness of 0.36 nm. No surface development in the form of hills, holes, or steps were observed. Despite a reduced substrate temperature for the last 3 min of growth, the growth mode for GaAs was still 2D (Stranski–Krastanov). The islands coalesced and formed narrow terraces, more visible in the contrasted SEM image than in AFM. Between 25 and 30 min of GaSb layer growth, the surface morphology had Gaussian-shaped hole regions about 70 nm deep with a half-width of approximately 200 nm, an average thickness of 190 nm, and overgrew to form undeveloped crystallization fronts. From 25 to 60 min of GaSb layer growth, there was a local minimum in surface roughness at the crystallization front, which temporarily varied between 25 and 60 min. Subsequently, the surface developed slightly with regular pyramids, maintaining a 2D growth mode with a local roughness maximum of 1.1 nm before stabilizing at 1.02 nm for a GaSb layer thickness of approximately 1400 nm. Increasing the GaSb buffer layer to 2 μm did not significantly improve surface roughness but reduced threading dislocations and peak half-width in the $\omega_{RC}$ scan to 190 arcsec. After 2.5 min of growth, numerous GaSb clusters with an average height of 36 nm formed on the GaAs surface, which fused after 5 min to form a mosaic-like surface with void spaces extending into the GaAs layer. After 15 min, there were oval holes in the 126 nm thick GaSb buffer layer with gentle slopes extending into the GaAs layer. After 25 min, the surface had numerous regular hole areas. After 60 min, the surface was free of hole areas and showed a 2D growth mode with regular pyramids and flowing atomic steps. After 300 min, the surface appeared as large terraced areas with flowing atomic steps and occasional threading dislocations. By comparing surface morphology results to pyrometric measurements, we estimated that the moment of obtaining an undeveloped surface occurred approximately 30 min into the GaSb buffer layer growth performed according to the M1 procedure, which resulted in a thickness of about 200–220 nm. The estimated moment of surface smoothing is indicated by the green arrow at the bottom of Fig. 4.

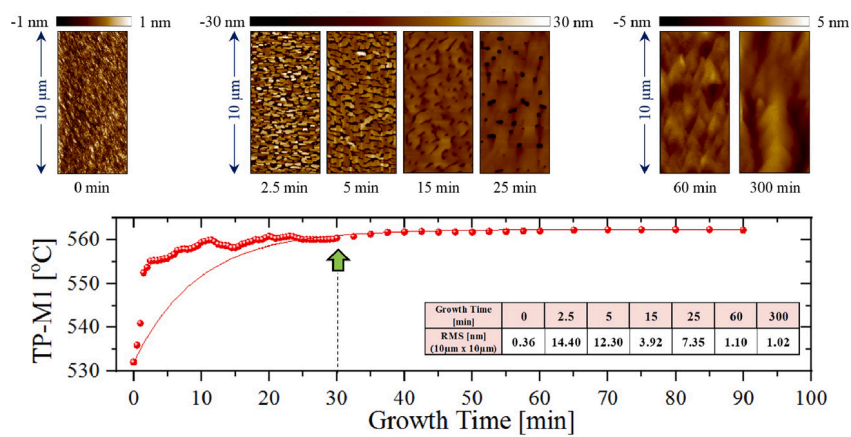

**Fig. 4.** Surface morphology of GaSb layers deposited according to the method described as M1 obtained from AFM measurements for selected GaSb layer growth times: 0, 2.5, 5, 15, 25, 60 and 300 min. The substrate temperature dependence as a function of the GaSb layer growth time performed according to the method described as M1 and the mathematical fit to the non-random measurement points are shown below. The inset contains the surface roughness of the GaSb layer obtained from AFM measurements from the (10 × 10) μm area for selected GaSb layer growth times: 0, 2.5, 5, 15, 25, 60 and 300 min. The green arrow indicates the predicted moment of overgrowth of the full GaSb layer without hole regions.

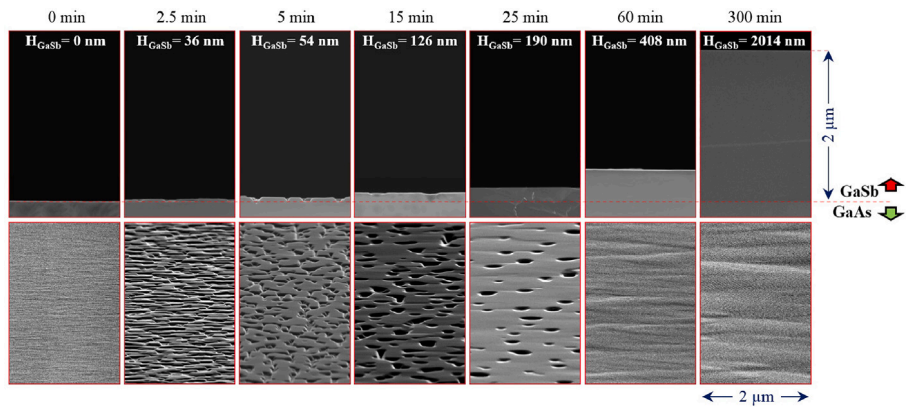

**Fig. 5.** SEM images from a cross-section of GaSb layers deposited according to the method described as M1 for selected GaSb layer growth times: 0, 2.5, 5, 15, 25, 60 and 300 min. The red dotted line indicates the location of the GaSb/GaAs layer interface. The $H_{GaSb}$ values placed at the top of the SEM image indicate the average height of the GaSb cluster or layer. Below is an SEM image from the surface of the samples tilted at 9°.

### 3.1.2. GaSb buffer layers deposited on GaAs (100) substrates by method labeled M2

The surface of the GaAs layer was wetted with an As flux for 15.5 min and then left unprotected for approximately 18 s before being cooled in an Sb flux for 348 s to a substrate temperature of $T_S = 532$ °C. The next step involved wetting the surface with a Sb flux for 552 s followed by cooling in a Sb flux. This resulted in a surface roughness of 0.29 nm. Fine oval undulations appeared on the GaAs surface, suggesting that the crystallization front was Ga-rich after interrupting the growth of the GaAs layer and reducing the As flux by three orders of magnitude while leaving the substrate unprotected for 18 s. Additionally, GaSb quantum objects with an average area of 26 450 nm² and a height of 2.14 nm were formed. These self-organized objects were attributed to GaSb nanoclusters resulting from various processes that occurred during the conversion of group V fluxes from As to Sb. Firstly, the absence of an As protective atmosphere led to the dissociation of As anions from the surface of the crystallization front, releasing Ga cations that diffused across the surface and formed GaSb nanoclusters upon contact with the Sb flux. Differences in surface tension between GaAs and GaSb caused the crystallization of GaSb in the form of nanoclusters or self-organized quantum dots. This effect was observed by another research group, which also showed that the instantaneous conversion of fluxes from Sb to As resulted in a transition from GaSb quantum dots to GaSb quantum rings due to the conversion of group V anions in GaSb dots from Sb to As and the diffusion of Sb anions to the base of the dot and the re-incorporation and formation of rings around partially distributed GaSb dots [17]. By comparing pyrometric measurements with surface morphology, a clear period was identified during which the substrate temperature linearly increased until reaching the undeveloped surface of the GaSb layer crystallization front for which the substrate temperature stabilized at $T_S = 562$ °C at approximately 42 min of growth and a GaSb layer thickness of about 280 nm. For GaSb layer thicknesses in the range of 280 to 400 nm, there was a local minimum of surface roughness at the crystallization front, which temporally varied from 42 to 60 min. Subsequently, the surface developed slightly maintaining a 2D growth with regular pyramids that reached a local roughness maximum of 1.24 nm before smoothing out to 0.81 nm for a GaSb buffer layer thickness of approximately 2030 nm. The GaSb buffer layer with a thickness of 2 μm had pronounced atomic steps and occasional threading dislocations, and the peak half-width in the $\omega_{RC}$ scan reached values of 214 arcsec. After approximately 2.5 min of growth, numerous GaSb clusters with an average height of

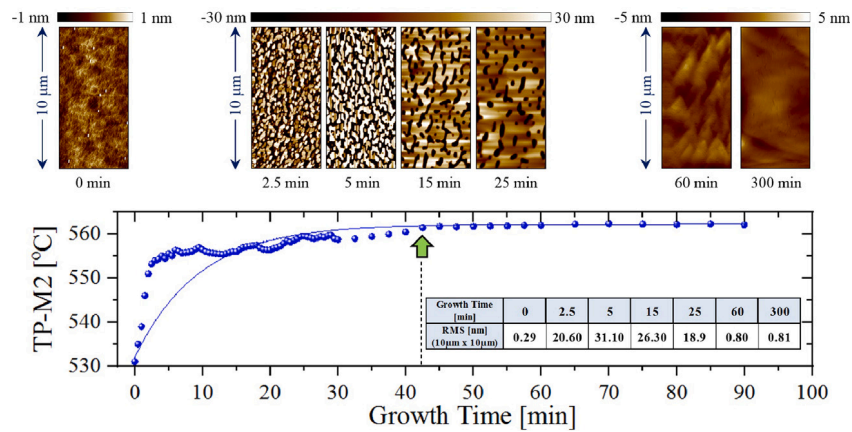

**Fig. 6.** Surface morphology of GaSb layers deposited according to the method described as M2 obtained from AFM measurements for selected GaSb layer growth times: 0, 2.5, 5, 15, 25, 60 and 300 min. The substrate temperature dependence as a function of the GaSb layer growth time performed according to the method described as M2 and the mathematical fit to the non-random measurement points are shown below. The inset contains the surface roughness of the GaSb layer obtained from AFM measurements from the (10 × 10) μm area for selected GaSb layer growth times: 0, 2.5, 5, 15, 25, 60 and 300 min. The green arrow indicates the predicted moment of overgrowth of the full GaSb layer without hole regions.

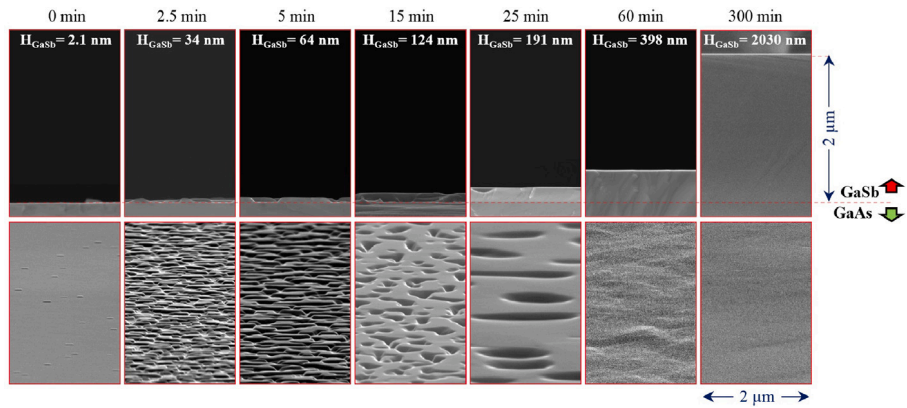

**Fig. 7.** SEM images from a cross-section of GaSb layers deposited according to the method described as M2 for selected GaSb layer growth times: 0, 2.5, 5, 15, 25, 60 and 300 min. The red dotted line indicates the location of the GaSb/GaAs layer interface. The $H_{GaSb}$ values placed at the top of the SEM image indicate the average height of the GaSb cluster or layer. Below is an SEM image from the surface of the samples tilted at 9°.

34 nm formed on the GaAs surface. After 5 min, the clusters grew taller and wider but remained separate from each other. After 15 min, the clusters grew together to form a mosaic surface with areas of void space extending into the GaAs layer. After 60 min, the surface was free of hole areas and indicated a 2D growth mode with regular pyramids and flowing atomic steps. By comparing the results of surface morphology with pyrometric measurements, it was estimated that the moment of undeveloped surface occurred at approximately 42 min of growth of the GaSb buffer layer performed according to the M2 procedure, which resulted in a layer thickness of about 280–300 nm. The estimated moment of surface smoothing is indicated by the green arrow in the lower part of Fig. 6.

### 3.1.3. GaSb buffer layers deposited on GaAs (100) substrates by method labeled M3

For the GaAs layer that was wetted for 1 min and cooled in the Sb flux, a surface roughness of 0.35 nm was achieved. No hills or holes were observed on the surface, nor were atomic steps visible. Despite the reduced substrate temperature during the final 3 min of growth for GaAs, the growth mode remained 2D (Stranski–Krastanov growth mode). The islands coalesced and formed narrow terraces that were better visible in the contrasted SEM image than on the AFM. By comparing the pyrometric measurements with the surface morphology, it was possible to estimate the period at which the randomness of temperature measurement disappeared. Between 15 and 18 min of growth for GaSb, the substrate temperature remained constant before increasing according to an exponential function and stabilizing at $T_S$ = 562 °C. During this time frame, between 15 and 18 min of GaSb layer growth, the surface morphology was rich in Gaussian-shaped hole regions approximately 79 nm deep with a half-width of around 220 nm at an average layer thickness of 117 nm, which overgrew to form the undeveloped surface of the crystallization front. For GaSb layer thicknesses ranging from 140 to 400 nm, a local minimum in the roughness of the crystallization front surface was observed, lasting from 18 to 60 min. The surface morphology then developed slightly while maintaining a 2D growth with regular pyramids, which reached a local roughness maximum of 1 nm before smoothing out to 0.57 nm for a GaSb buffer layer thickness of approximately 2066 nm. The half-width of the peak in the $\omega_{RC}$ scan for a GaSb buffer layer of 2066 nm thickness reached values of 176 arcsec. Based on the surface roughness trend shown in Fig. 10, an increase in the GaSb buffer layer thickness below 2 micrometers should further reduce the surface roughness. After 2.5 min of growth for the Be-doped GaSb layer, a mosaic-like growth of intergrown GaSb:Be clusters with an average height of 40 nm formed on the GaAs surface. After 5 min, the GaSb:Be mosaic layer reached a height of 60 nm with areas of void space reaching the GaAs layer. After 15 min, the 117 nm thick GaSb buffer layer had numerous oval holes with an average depth of 79 nm and gentle slopes. After 25 min, the surface was free of hole regions and exhibited a 2D growth mode with regular pyramids of flowing atomic steps. After 60 min of GaSb:Be layer growth, pyramids with flowing atomic steps were still visible on the surface. Following 300 min of growth, pyramids were no longer observable on the surface, only flowing atomic steps remained. By comparing the results of the surface morphology with the pyrometric measurements, it was estimated that the moment of obtaining an undeveloped surface occurred at approximately 18 min of GaSb:Be buffer layer growth performed according to the M3 procedure, resulting in a layer thickness of approximately 140–160 nm. This estimated time of surface smoothing is indicated by the green arrow in the lower part of Fig. 8.

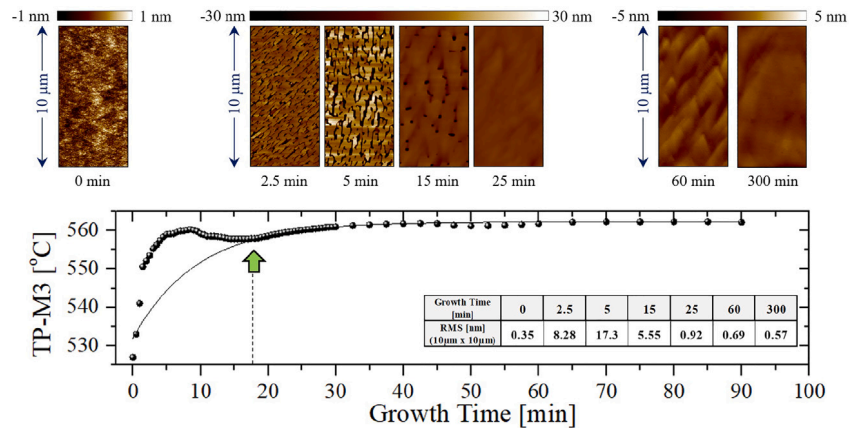

**Fig. 8.** Surface morphology of GaSb layers deposited according to the method described as M3 obtained from AFM measurements for selected GaSb layer growth times: 0, 2.5, 5, 15, 25, 60 and 300 min. The substrate temperature dependence as a function of the GaSb layer growth time performed according to the method described as M3 and the mathematical fit to the non-random measurement points are shown below. The inset contains the surface roughness of the GaSb layer obtained from AFM measurements from the (10 × 10) μm area for selected GaSb layer growth times: 0, 2.5, 5, 15, 25, 60 and 300 min. The green arrow indicates the predicted moment of overgrowth of the full GaSb layer without hole regions.

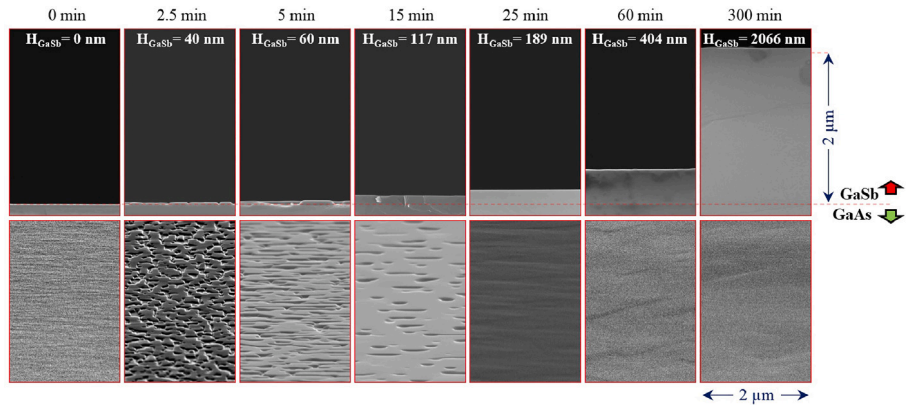

**Fig. 9.** SEM images from a cross-section of GaSb layers deposited according to the method described as M3 for selected GaSb layer growth times: 0, 2.5, 5, 15, 25, 60 and 300 min. The red dotted line indicates the location of the GaSb/GaAs layer interface. The $H_{GaSb}$ values placed at the top of the SEM image indicate the average height of the GaSb cluster or layer. Below is an SEM image from the surface of the samples tilted at 9°.

### 3.2. Comparison of the above M1, M2 and M3 methods for the deposition of GaSb layers on GaAs substrates using MBE technology

The behavior of the surface morphology followed a similar pattern for all three methods, initially exhibiting an increase in roughness as GaSb clusters formed before decreasing after forming a solid layer, reaching a minimum value and then slightly increasing with the emergence of regular pyramid-shaped structures. The half-widths (FWHM) of the GaSb peaks in the $\omega_{RC}$ scan for GaSb buffer layers with an approximate thickness of 2 μm were 214 arcsec, 190 arcsec, and 176 arcsec for methods M2, M1, and M3, respectively. The surface roughnesses (RMS) were 1.02 nm, 0.81 nm, and 0.57 nm for methods

M1, M2, and M3, respectively. According to Fig. 10 b inset, further smoothing of the GaSb surfaces is expected if growth continues using method M3 beyond a thickness of 2 μm. In Figs. 11 a-c, which depict double enlarged AFM scan areas of approximately (10 × 10) μm for GaSb layers with approximate thicknesses of 2 μm made by each of the three methods, defects visible on the surface are marked using a red circle for methods M1 and M2 while threading dislocations were observed in method M3 at a layer thickness of approximately 1 μm but not at a thickness of approximately 2 μm.

The improvement of the GaSb layer parameters compared to the methods marked as M1 and M2 is attributed to two factors: Be doping and substrate subcooling by 5 °C in the initial phase of GaSb growth. Firstly the presence of Be dopant in the initial phase of the GaSb layer crystallization effectively reduces Ga mobility along the surface along the [110] and [1-10] directions before locating GaSb clusters on the surface through the impurity-induced mixing effect of disordered layers [18]. The extended clusters merge to generate threading dislocations TDs defects after the interaction of misfit dislocations 60° MDs defects present in the leading edges of individual GaSb clusters with misfit dislocations 90° MDs defects present in the GaSb layer [19]. When the coalescence areas reach upper parts of the clusters, new 60° MDs are formed at the coalescence zones. The presence of Be hardens the GaSb alloy by solid solution, *i.e.*, limits the dislocation mobility. These new 60° MDs formed on coalescence zones, are pinned by the lattice distortions produced by the Be dopants. In this situation, the dislocations cannot glide down to the GaSb/GaAs interface where interactions with the 90° MDs produce more TDs that propagate from the interface plane to the top surface, but pushed by the GaSb layer lattice strain and their interaction with the 90° MD is avoided. The total GaAs buffer layer surface coverage with GaSb or GaSb:Be material after 2.5 min of growth 87.5%, 76.6%, and 65.8% for M3, M1, and M2 methods, respectively. It should be pointed out, that in M3 method, there was not enough time to stabilize the substrate temperature after cooling the main heater from $T_H = 561$ °C to $T_H = 445$ °C. This may had a direct impact on the rapid formation of the mosaic-like surface just after 2.5 min of GaSb buffer layer growth. The rapid cooling with a ramp of approximately 38 °C/min resulted in substrate overcooling by about 20 °C. It took at least 3 min for the substrate to reach a stable temperature of $T_S = 532$ °C from the time it reached the set temperature on the main heater. In method M3, the period for stabilization was limited to 61 s, and thus, GaSb layer growth commenced on a thermally unstabilized substrate. The pyrometer indicated a substrate temperature of $T_S = 527$ °C just before the start of GaSb layer growth for this one method only. Substrate subcooling in the initial growth phase influences the type of dislocations that form. 90° MDs are observed at low growth temperatures (520 °C), while 60° ones appear at high temperatures (560 °C) [20]. Lowering the substrate temperature in the initial phase of GaSb growth reduces Ga's mobility across the surface and its tendency to build up on top of the GaSb cluster making the clusters lower and wider and their coalescence faster.

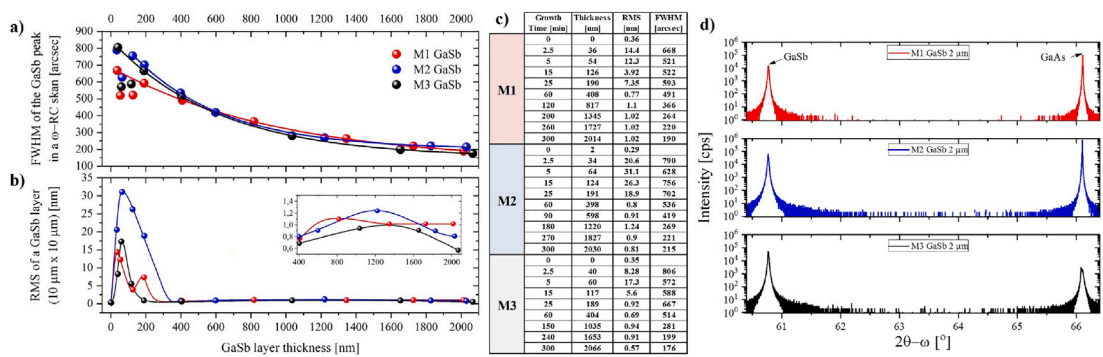

**Fig. 10.** (a) Dependence of GaSb peak half-width FWHM obtained from $\omega_{RC}$ scan on cluster or GaSb layer thickness obtained from SEM. (b) Dependence of surface roughness RMS of GaSb layer obtained from AFM (10 × 10) μm scan on cluster or GaSb layer thickness obtained from SEM. The inset contains the area for GaSb layer thicknesses from 400 to 2000 nm with an enlarged scale. The solid lines are the proposed momentum lines for each method. (c) Table compiling the growth time, GaSb cluster or layer thickness, RMS surface roughness and FWHM GaSb peak half-width for the $\omega_{RC}$ scan for each method. (d) HR-XRD results for a 2$\Theta$-$\omega$ scan of GaSb layers of approximately 2 μm thickness performed according to procedures labeled M1, M2 and M3, respectively..

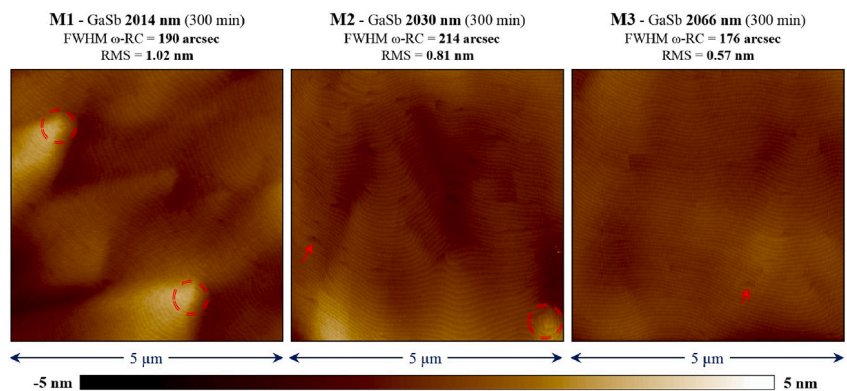

**Fig. 11.** AFM measurements from the (5 × 5) μm area for approximately 2 μm GaSb buffer layers made according to 3 procedures labeled M1, M2 and M3. Above the figures are information on layer thickness from the SEM, half-width FWHM for the GaSb peak from the $\omega_{RC}$ scan and surface roughness RMS from the AFM for the (5 × 5) μm area. Red circles and arrows indicate visible defects on the surface.

## 4. Conclusions

During the initial stages of GaSb buffer layer crystallization on GaAs substrates, using three distinct methods identified as M1, M2, and M3, the evolution of surface morphology was analyzed to examine its behavior. Among these techniques, method M3 produced the most favorable results for a GaSb buffer layer with a thickness exceeding 2 micrometers. Our observations indicate that GaSb buffer layers with an approximate thickness of 2 μm should be compared as at this thickness, the layers stabilize and the surface morphology is beyond the local roughness minimum. The quality of the GaSb buffer surface plays a crucial role in determining the quality of the subsequent layers that form the entire T2SL device. Specifically, from the perspective of application, the surface quality is critical for understanding defects in the final device, such as the amount of dark current.

**CRediT authorship contribution statement**

**Dawid Jarosz:** Writing – review & editing, Writing – original draft, Visualization, Validation, Supervision, Methodology, Investigation, Formal analysis, Data curation, Conceptualization. **Ewa Bobko:** Writing – review & editing, Data curation. **Marcin Stachowicz:** Writing – original draft, Formal analysis. **Ewa Przeździecka:** Writing – original draft, Formal analysis. **Piotr Krzemiński:** Data curation. **Marta Ruszała:** Data curation. **Anna Juś:** Data curation. **Małgorzata Trzyna-Sowa:** Data curation. **Kinga Maś:** Data curation. **Renata Wojnarowska-Nowak:** Data curation. **Oskar Nowak:** Formal analysis. **Daria Gudyka:** Formal analysis. **Brajan Tabor:** Formal analysis. **Michał Marchewka:** Writing – review & editing, Writing – original draft, Formal analysis.

**Declaration of competing interest**

The authors declare that they have no known competing financial interests or personal relationships that could have appeared to influence the work reported in this paper.

**Data availability**

Data will be made available on request.

**Acknowledgments**

This research was funded by the National Centre for Research and Development (NCBR), Poland, under project No. POIR.04.01.04-00-0123/17 and under project No. SKN/SP/601031/2024.